\documentstyle[aps,prl,multicol,floats,epsf]{revtex}

\newcommand{\bi}{\begin{itemize}\setlength{\itemsep}{0pt}}
\newcommand{\ei}{\end{itemize}}
\newcommand{\be}{\begin{equation}}
\newcommand{\ee}{\end{equation}}
\newcommand{\bea}{\begin{eqnarray}}
\newcommand{\eea}{\end{eqnarray}}
\def\(#1){(\ref{#1})}

\newcommand{\bfig}[3]{\begin{figure}\vspace*{#2}\begin{center}\leavevmode\epsfxsize #1\epsfbox{#3.eps}}
\newcommand{\efig}[2]{\end{center}\vspace*{-0.2cm}\caption{#2\label{fig:#1}}
\end{figure}}

\renewcommand{\l}{l}
\renewcommand{\k}{k}
\renewcommand{\propto}{\sim}

\newcommand{\E}{E}
\newcommand{\Emax}{E_{\rm max}}
\newcommand{\ommin}{\om_{\rm min}}

\newcommand{\Gn}{\Gamma_0}
\newcommand{\G}{\Gamma}
\newcommand{\lc}{{l_{\rm y}}}
\newcommand{\rh}{\rho}
\newcommand{\gam}{\gamma}

\newcommand{\gamdot}{{\dot{\gamma}}}
\newcommand{\eq}{_{\rm eq}}
\newcommand{\sst}{_{\rm ss}}
\newcommand{\g}{_{\rm g}}
\newcommand{\sig}{\sigma}
\newcommand{\sigy}{\sigma_{\rm y}}
\newcommand{\const}{{\rm const}}

\newcommand{\zz}{z}

\newcommand{\visc}{\eta}

\newcommand{\om}{\omega}
\newcommand{\Gcomp}{G^*}

\newcommand{\half}{{1\over 2}}
\newcommand{\deriv}[1]{{\partial\over\partial#1}}
\newcommand{\lav}{\left\langle}
\newcommand{\rav}{\right\rangle}

\newcommand{\ie}{{i.e.},}
\newcommand{\eg}{{e.g.},}
\newcommand{\rhs}{r.h.s.}

\draft
\begin{document}

\title{Rheology of Soft Glassy Materials}
\author{Peter Sollich$^1$\cite{fellow_email}, 
Fran\c cois Lequeux$^2$, Pascal
H\'ebraud$^2$, Michael E Cates$^1$}
\address{$^1$Department of Physics and Astronomy, University of Edinburgh,
Edinburgh EH9 3JZ, U.K.\\
$^2$Laboratoire d'Ultrasons et de Dynamique des Fluides Complexes,
 4 rue Blaise Pascal, 67070
Strasbourg Cedex, France}

\maketitle

\begin{abstract}
We attribute similarities in the rheology of many soft
materials (foams, emulsions, slurries, etc.)  to the shared features of
structural disorder and metastability.  A generic model for
the mesoscopic dynamics of ``soft glassy matter'' is
introduced, with interactions represented by a mean-field noise
temperature $x$.
We find power law fluid behavior either with
($x<1$) or without ($1<x<2$) a yield stress.  For $1<x<2$, both
storage and loss modulus vary with frequency as $\omega^{x-1}$,
becoming flat near a glass transition ($x=1$).  Values
of $x\approx 1$ may result from marginal dynamics as seen in some
spin glass models.
\end{abstract}

\pacs{
Submitted to {\em Physical
Review Letters}. 
}

\vspace{-0.9\baselineskip}
\begin{multicols}{2}
Many soft materials, such as foams, emulsions, pastes and slurries, have
intriguing rheological properties. Experimentally, there is
a well-developed phenomenology for such systems: their nonlinear flow behavior
is often fit to the form  $\sigma = A + B \dot \gamma^n$
where $\sigma$ is shear stress and $\dot\gamma$ strain rate.
This is the Herschel-Bulkeley equation~\cite{Holdsworth93,Dickinson92};
or (for $A=0$) the ``power-law
fluid''~\cite{Holdsworth93,Dickinson92,BarHutWal89}.
For the same materials, linear or quasi-linear viscoelastic
mesurements often reveal storage and loss
moduli $G'(\omega)$,
$G''(\omega)$ in nearly constant ratio
($G''/G'$  is usually about 0.1) with a frequency dependence that is
either a weak power law (clay slurries, paints, microgels) or
negligible (tomato paste, dense emulsions, dense multilayer vesicles,
colloidal glasses)
\cite{MacMarSmeZha94,KetPruGra88,KhaSchneArm88,MasBibWei95,PanRouVuiLuCat96,HofRau93,MasWei95}.
This behavior persists down to the
lowest accessible frequencies (about
$10^{-3}$--1 Hz depending on the system), in apparent contradiction to
linear response theory~\cite{BuzLuCat95}, which requires that $G''(\om)$ should be an
odd function of $\om$.

That similar anomalous rheology should be seen in such a wide range of soft
materials
suggests a common cause. Indeed, the frequency dependence indicated above points
strongly to the generic presence of slow ``glassy'' dynamics persisting to
arbitrarily
small frequencies. This feature is found in several other
contexts~\cite{MonBou96,BouDea95,Bouchaud92}, such as
elastic manifold dynamics in random
media~\cite{VinMarChe96,LeDouVin95}.
The latter is suggestive of rheology:
charge density waves, vortices, contact lines, etc. can
``flow'' in response to an imposed ``stress''.
In this Letter we argue that glassy dynamics is a natural consequence of two
properties shared by all the soft materials mentioned above:
{\em structural disorder} and  {\em metastability}. In such materials,
thermal motion alone is not enough to achieve complete structural
relaxation. The system has to
cross energy barriers (for example those associated with rearrangement of
droplets in an emulsion) that are very large compared to
typical thermal energies. Therefore the system adopts a disordered, metastable
configuration even when
(as in a monodisperse emulsion or foam) the state of least
free energy would be ordered~\cite{emulsion_inherent_metastability}.
While the importance
of disorder has been noted before for specific
systems~\cite{MasBibWei95,BuzLuCat95,WeaFor94,LacGreLevMasWei96,OkuKaw95,Durian95},
we feel that its unifying role in rheological modelling has not
been appreciated.

To test these ideas, we
construct a minimal ``generic model'' for soft glassy matter.
For simplicity, we ignore tensorial aspects, restricting our analysis
to simple shear strains. Consider first
the behavior of a foam or dense emulsion under shear. We focus on a
{\em mesoscopic}
region, large enough for a local strain variable $\l$ to be defined, but
small enough
for this to be approximately uniform within the region, whose size
we choose as the unit of length.
As the system is sheared,
droplets in this region will first deform
elastically from a local equilibrium configuration, giving rise to
a stored elastic energy (due to surface tension, in this
example~\cite{WeaFor94}).
This continues
up to a yield point, characterized by a strain $\lc$, whereupon the
droplets rearrange to new positions in which they are less deformed,
thus relaxing stress.
The mesoscopic strain $\l$
{\em measured from the nearest equilibrium position} (\ie\ the one which can be
reached by purely elastic deformation) therefore executes a saw-tooth
motion as the macroscopic strain $\gam$ is increased~\cite{sawtooth}.
Neglecting nonlinearities
before yielding, the local shear stress is given by $\k\l$, with $\k$
an elastic constant; the yield point defines a maximal elastic
energy $\E=\half\k\lc^2$.
A similar description obviously extends to many others of the
soft materials discussed above.

We now ascribe to each mesoscopic region
not only its own strain variable $\l$, but also its own maximal
yield elastic energy, $\E>0$. We model the effects of
structural disorder by assuming a {\em distribution} of such yield energies
$\E$, rather
than a single value common to all regions.  The state of a macroscopic
sample is then
characterized by a probability distribution $P(\l,\E;t)$.
We propose the following dynamics for the
time evolution of $P$:
\be
\deriv{t}P =  -
\gamdot\deriv{\l}P-\Gn e^{-(\E-\half\k\l^2)/x}\,P + \G(t)\,\rh(\E)\delta(\l)
\label{basic}
\ee
The {first} term on the \rhs\ arises from the elastic deformation of the
regions. This embodies a mean
field assumption, that between successive local yield events, changes in
local strain follow those
of the macroscopic deformation: $\dot{\l}=\dot{\gamma}$. Note, however,
that due to stochas-
tic yielding events the stress $\k\l$ is spatially inhomoge-
\end{multicols}
\newpage
\twocolumn\noindent
neous (as is the local strain $\l$).
The macroscopic stress
is defined as an
average over regions
\be
\sig(t)=\k\lav\l\rav\equiv\k \int\!\,\l\,P(\l,\E;t)\,d\l\, d\E \ .
\label{stress_def}
\ee
The {second} term on the \rhs\ of~\(basic) describes the yielding
of our mesoscopic regions. We have written the yielding rate as the
product of an
``attempt frequency'' $\Gn$, and an exponential
probability for activation over an energy barrier $\E-\half\k\l^2$ (the
excess of the yield energy over that stored elastically).
However, the resemblance to thermal activation is formal:
we expect these ``activated'' yield processes to arise primarily by coupling to
structural rearrangements elsewhere in the system. In a mean-field spirit,
all such interactions between regions
are subsumed into an
effective ``noise temperature'', $x$. We first regard $\Gn,x$ as arbitrary
constants,
but later discuss their meaning and their
possible dependences on other quantities.

Finally, the {third} term on the \rhs\ of~\(basic) describes the
relaxation of regions to new local equilibrium positions after yielding,
which we treat
as effectively instantaneous.
The first factor in this term is simply the total yielding rate
$\G(t)$ $=$ $\Gn$ $\lav \exp[-(\E-\half\k\l^2)/x]\rav_P$.
The remaining two factors incorporate
further mean-field assumptions, as follows.
First, the yield energy $\E$ for distortions about any
equilibrium configuration is uncorrelated with the previous one for this
region; it is
drawn randomly from the prior distribution (``density of states'')
$\rho(\E)$ which we assume to be time-independent.
Second, immediately after yielding, a region always finds itself in
a completely unstressed state of local equilibrium with $\l=0$
(hence the Dirac delta function, $\delta(\l)$). This latter simplification is
not essential, as shown elsewhere~\cite{long_El}.

In the absence of flow ($\gam(t)=0$), the model~\(basic) describes
activated hopping between ``traps'' of depth $\E'=\E-\half\k\l^2$ with
density $\rho(\E')$. This corresponds to Bouchaud's model for glassy
dynamics~\cite{MonBou96,BouDea95,Bouchaud92}, whose predictions
we briefly recall.
For high (noise) temperatures $x$ the system evolves towards
the Boltzmann distribution $P\eq(\E')\propto \rh(\E')\exp(E'/x)$. As $x$
is lowered, this distribution may cease to be
normalizable, leading to a glass transition at
$x\g^{-1}=-\lim_{E\to\infty}(\partial/\partial\E) \ln\rh(\E)$.
For $x<x\g$, no equilibrium state exists, and
the system shows ``weak ergodicity breaking''
and various aging phenomena.
A finite value of $x\g$ implies
an
exponential tail in the density of states,
$\rho\sim\exp(-E/x\g)$, which corresponds to a Gaussian
distribution of yield strains $\lc=(2\E/\k)^{1/2}$.
\bfig{9cm}{-0.8cm}{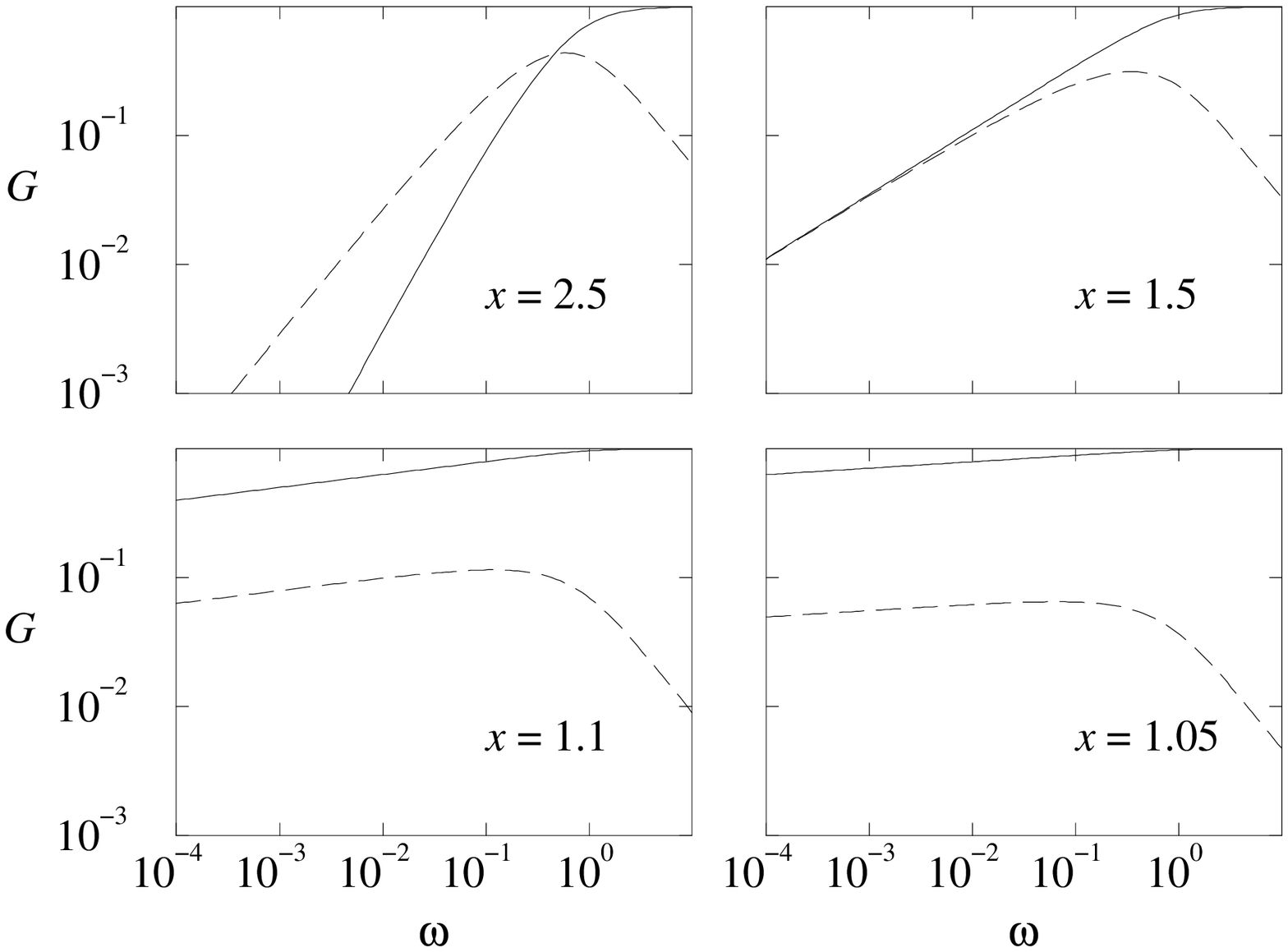}
\efig{linear_moduli}{Linear moduli $G'$ (solid line) and $G''$
(dashed) vs
frequency $\om$ at various noise temperatures.}

A major attraction of the model defined
by~(\ref{basic}) and~(\ref{stress_def}) is that an exact constitutive
equation, relating the stress $\sigma(t)$ to the strain-rate history
$[\gamdot(t'<t)]$,
can be obtained~\cite{long_El}.  Since this is quite complicated,
we restrict ourselves here to two
standard rheological tests, for which the full form is not required. We use
non-dimensional units for time and
energy by setting $\Gn=x\g=1$; we also rescale our strain variables
($\l,\gamma$) so that $\k=1$. In these units,
$\rh(\E)=\exp[-\E(1+f(\E))]$ with $f(\E)\to
0$ for $\E\to\infty$. Up to sub-power-law factors such as logarithms,
all power laws reported below are valid for any $f(\E)$; numerical
examples use $f\equiv 0$. Analytical 
and numerical support for our results
will be detailed elsewhere~\cite{long_El}.

Consider first the complex dynamic shear modulus
$\Gcomp(\omega)=G'+iG''$, which describes the stress response to small shear
strain perturbations around the equilibrium state. As such, it is well
defined (\ie\ time-independent) only above the glass
transition, $x>1$.  Expanding~\(basic) to first order in the amplitude
$\gam$ of an oscillatory strain $\gam(t)=\gam\cos\om t$, we find
$
\Gcomp(\om)=\lav{i\om\tau/(i\om\tau+1)}\rav\eq
$. This
corresponds to a distribution of Maxwell modes
whose spectrum of relaxation times $\tau=\exp(\E/x)$
is given by the equilibrium distribution
$P\eq(\E)\propto\exp(\E/x)\rho(\E)$. The relaxation time spectrum thus
exhibits
power law behavior for large $\tau$: $P(\tau)\sim \tau^{-x}$.
This leads to power laws for $\Gcomp$ in the low frequency range
(Fig.~\ref{fig:linear_moduli}):
\be
\begin{array}{lcllcll}
G'' & \propto &\omega \      &\mbox{for $2<x$}, \quad
    & \propto &\omega^{x-1}\ &\mbox{for $1<x<2$} \\
G' & \propto &\omega^2 \     &\mbox{for $3<x$}, \quad
    & \propto &\omega^{x-1}\ &\mbox{for $1<x<3$}
\end{array}
\ee
For $x>3$ the system is Maxwell-like at low frequencies, whereas
for $2<x<3$ there is an anomalous power law in the elastic modulus.
Most interesting is the regime $1<x<2$,
where $G'$ and $G''$ have constant ratio;
both vary as $\omega^{x-1}$.
Behavior like this is observed in a number of soft
materials \cite{MacMarSmeZha94,KetPruGra88,KhaSchneArm88,MasBibWei95,MasWei95}.
Moreover, the frequency exponent approaches zero as $x\to 1$, resulting in
essentially constant values of $G''$ and $G'$, as reported
in dense emulsions, foams,
and onion phases~\cite{KhaSchneArm88,MasBibWei95,PanRouVuiLuCat96}. Note,
however, that the ratio
$G''/G'\propto x-1$ becomes small
as the glass transition is
approached. This increasing dominance of the elastic response $G'$
prefigures the onset of a yield stress for $x<1$
(discussed below)~\cite{crossover}.
If a high energy cutoff $\Emax$ is imposed on $\rho(E)$ (giving
an upper limit on local yield strains), the above results remain valid down to
$\ommin=\exp(-\Emax/x)$. Well-defined
equilibrium values of the linear moduli then exist also for $x<1$;
one still finds $G''\propto \om^{x-1}$ for
$\ommin\ll\om\ll 1 $. For $x$ just below
$x\g=1$, a log-log plot of
$G''(\om)$ therefore exhibits a small {\em negative} slope
(whereas $G'$ is constant).
This may
again be compatible
with recent experimental data
~\cite{MasBibWei95,PanRouVuiLuCat96,HofRau93,MasWei95}.

\bfig{9cm}{-1cm}{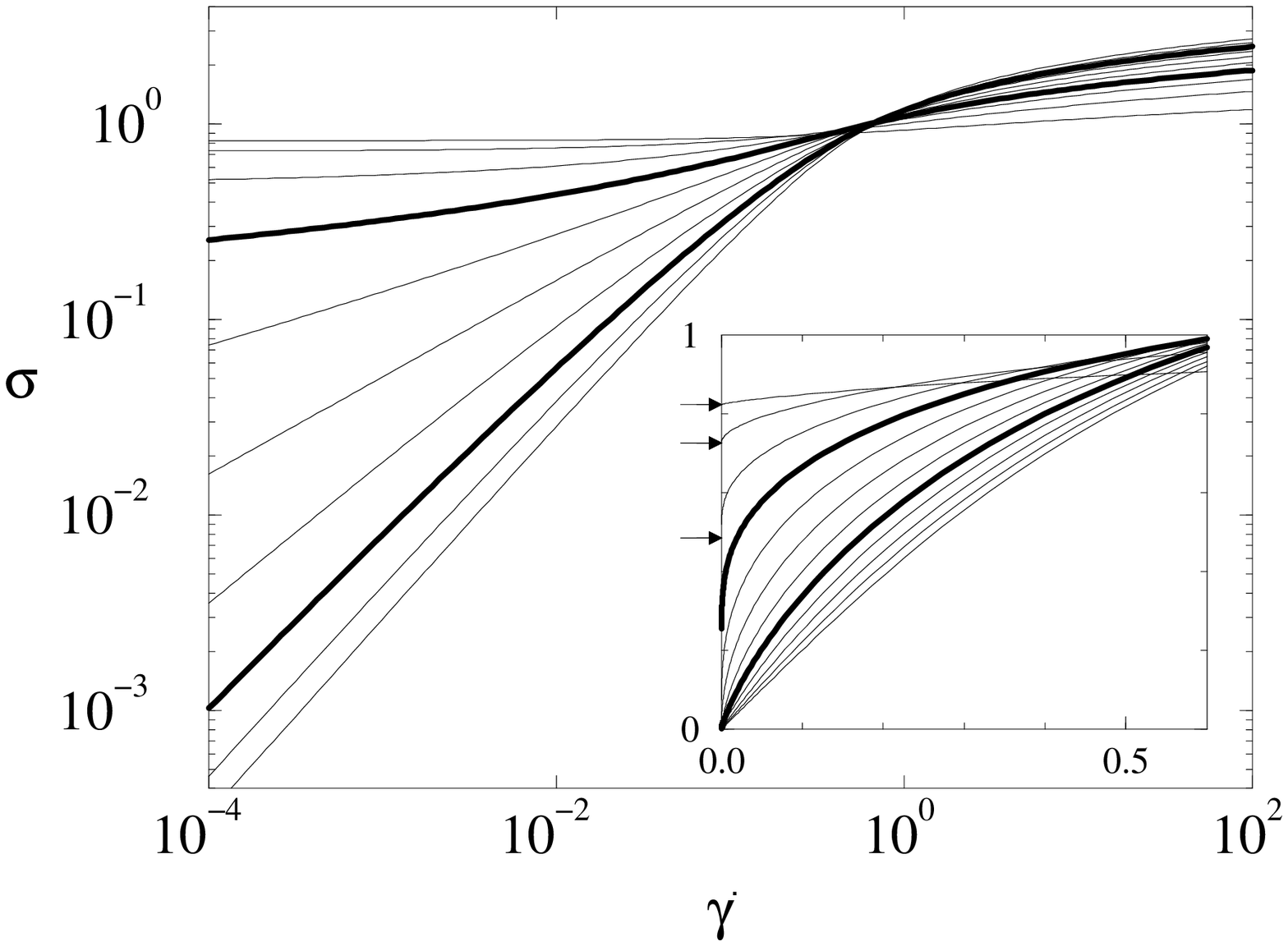}
\efig{flow_curves}{Shear stress $\sig$ vs shear rate $\gamdot$, for
$x=0.25$, 0.5, $\ldots$, 2.5 (top to bottom on left); $x=1,2$ are shown in
bold. Inset: small $\gamdot$ behavior, with yield stresses for $x<1$
shown by arrows.}
We now turn to the case of steady shear flow,
$\gamdot=\const$, for which the
steady state distribution $P\sst(\l,\E)$ can be
obtained analytically.
After integrating over $\E$, one finds $P\sst(\l) \propto
\Theta(l)g(\zz(l))$ with
\[
\zz(\l)={1\over\gamdot}\int_0^\l\!e^{\gam^2/2x}d\gam\quad
g(\zz)=
\int\! \rh(\E) \exp(-\zz e^{-\E/x})\,dE
\]
In the large $z$ limit, $g(\zz)\sim \zz^{-x}$.
Figure~\ref{fig:flow_curves} shows that
for large shear rates, $\gamdot\geq 1$, $\sig$ increases
very slowly for all $x$ ($\sig\sim(x\ln\gamdot)^{1/2}$).
More interesting is the small
$\gamdot$ behavior, where we find three regimes:
(i) For $x>2$, the system is Newtonian,
$\sig=\visc\gamdot$. The viscosity is simply the average relaxation time
$\visc=\lav \exp(\E/x) \rav\eq =\lav\tau\rav\eq$
taken over the equilibrium distribution of
energies, $P\eq(\E)\propto \exp(\E/x)\rho(\E)$. Hence
$\visc\propto\lav\exp(2\E/x)\rav_\rho$, which diverges at $x=2$. (ii)
For $1<x<2$
one finds power law fluid
behavior, $\sig\propto \gamdot^{x-1}$.
(iii) For $x<1$, the system shows a yield stress:
$\sig(\gamdot\to 0)=\sigy>0$. (This has a
linear onset near the glass
transition, $\sigy\propto 1-x$.)
Beyond yield, the stress again increases as a
power law of shear rate, $\sig-\sigy\propto \gamdot^{1-x}$ (for $\gamdot\ll 1$).
The behavior of our model in regimes (ii) and (iii) therefore matches
respectively
the power-law fluid ~\cite{Holdsworth93,Dickinson92,BarHutWal89} and
Herschel-Bulkeley~\cite{Holdsworth93,Dickinson92} 
scenarios as used to fit the nonlinear
rheology of pastes,
emulsions, slurries, etc.

We now speculate on the origin and magnitude of
the ``attempt frequency'' $\Gn$ and the ``noise temperature'' $x$.
First note that the parameter $\Gn$ is the only source of a
characteristic timescale (chosen as the time unit above).
We have approximated it by a constant value: $\Gn(\gamdot) = \Gn(0)$.
One possibility is that
the intrinsic rate constant  $\Gn$
arises from {\em true} thermal processes.
If so it can be estimated
as $\Gamma_{\rm loc}k_{\rm B}TP\eq(0)Q$ with $\Gamma_{\rm loc}$
a local diffusive attempt rate (for 1 $\mu$m emulsions this might be $0.01$s);
$k_{\rm B}TP\eq(0)$ is the (small) fraction of regions in which true
thermal activation can surmount the yield barrier. The factor
$Q$ denotes the number of neighboring regions perturbed as a result of
one such thermal event. A more detailed analysis (involving
an extension to our model \cite{long_El}) then shows
that $k_{\rm B}TP\eq(0)Q$ must be large enough (at least of order unity)
to 
avoid depletion of the low energy part ($E\leq k_{\rm B}T$) part of
the barrier distribution.
This mechanism may arise in systems (such as foams) in which
one local rearrangement can trigger a long sequence of
others~\cite{OkuKaw95,Durian95}.
If so, the resulting intrinsic rate
$\Gn\sim \Gamma_{\rm loc}$ provides a plausible rheological timescale. (If $Q$
is too small,
$\Gn$ will instead be of order $\Gamma_{\rm loc}e^{-\bar{E}/k_{\rm B}T}$,
which for typical barrier energies $\bar{E}=\lav\E\rav_\rho$ is 
unfeasibly slow.)

We emphasize, however, that $\Gn$ may be strongly system dependent,
and any specific
interpretation of it remains speculative.
Nonetheless we may view the activation factor in Eq.(1)
as the probability
that a perturbative ``kick'' to a given mesoscopic region (from events
elsewhere) causes
it to yield. We believe this activation factor should be primarily
geometric in origin and hence depend on
the disorder, but not on any intrinsic
energy scale. Accordingly (in our units) $x$ values generically
of order unity can be expected. We argue next that $x$ values {\em
close to} unity may be normal.

Consider first a steady shear experiment.
For soft metastable materials, the rheological
properties of a sample freshly loaded into a rheometer are usually
not reproducible; they become so only after a period of shearing
to eliminate memory of the loading
procedure. In the process of loading one expects a large degree of
disorder to be introduced;
the initial dynamics under flow should therefore involve a high noise
temperature
$x\gg 1$. As the sample approaches the steady state,
the flow will (in many cases)
tend to eliminate much of this macroscopic disorder~\cite{WeaBolHerAre92}
so that $x$ will decrease.
But, as this occurs, the noise-activated processes will slow down;
as $x\to 1$, they become negligible. Assuming that,
in their absence, the disorder cannot be
reduced further, $x$ is then ``pinned'' at
a steady-state value at or close to the glass transition.
This scenario, although extremely speculative,
is strongly reminiscent of the ``marginal dynamics'' seen in some mean-field spin
glass models~\cite{CugKur93note}.

There remains several ambiguities within this picture, for example whether
the steady state value of $x$ should depend on $\gamdot$; if it does so strongly,
our results for steady flow curves will of course be changed.
If a steady flow is stopped and a linear viscoelastic spectrum measured,
the behavior observed should presumably pertain to the $x$ characterizing
the preceding steady flow (assuming that $x$ reflects structure only).
But unless the strain amplitude is extremely small the $x$-value obtained
in steady state
could be affected by the oscillatory flow itself~\cite{footflat}.

Also uncertain is to what extent a steady energy input
is needed to sustain the nonlinear dynamics. Although not represented in
the model,
a small finite strain rate amplitude might be needed
to balance the gradual dissipation of energy
in yield events.
In its absence, one might expect the sample to
show aging (i.e., $P(l,E;t)$ nonstationary in time). Within the model, aging
in fact only occurs for $x<1$~\cite{Bouchaud92} (the regime for which we
predict a yield stress).
Conversely we saw above 
that, even in this regime,
for finite $\dot \gamma$ a well-defined steady state distribution is recovered:
{\em flow interrupts aging}~\cite{BouDea95}. This can be understood by considering the
distribution of
energies. Without flow, one obtains a Boltzmann distribution
$P(\E)\propto\rh(\E)e^{\E/x}$ up to (for $x<1$) a cutoff which
shifts to higher and higher energies as the system
ages~\protect\cite{MonBou96}. This cutoff, and hence the most long-lived traps
visited (which have a lifetime comparable to the age of the system),
dominate the aging behavior~\protect\cite{Bouchaud92}. The presence of flow
leads to a steady state value of this cutoff of $\E\sim
x\ln(\gamdot^{-1}x^{1/2})$, while for higher energies one has
$P\sst(\E)\propto\rh(\E) \E^{1/2}$. Hence flow prevents
regions from getting stuck in progressively deeper traps
and the aging process is truncated
after a finite time.

We are currently investigating more
complicated nonlinear strain histories~\cite{long_El}.
In future work, explicit spatial
structure and interactions between regions must be added
so as to understand better the mutual dynamical evolution
of the attempt rate, the effective noise temperature and the disorder.
One issue concerns the relative importance of localized
~\cite{LacGreLevMasWei96,HebMunLeqPin,LiuRamMasGanWei96,HutWeaBol95,GopDur95,DurWeiPin91,EarWil96}
versus avalanche-like~\cite{OkuKaw95,Durian95} events in the
relaxation of stress.

The authors are indebted to  J.-P.~Bouchaud for various seminal
suggestions. They also thank him, M.~O.~Robbins
and D.~Weaire for helpful discussions, and the Newton Institute,
Cambridge, for hospitality.
PS is a Royal Society Dorothy
Hodgkin Research Fellow.


\begin{thebibliography}{10}
\bibitem{emulsion_inherent_metastability}
Soft systems may also be intrinsically metastable in a more drastic sense (for
  example, with respect to coalescence in emulsions) -- we ignore this here.
\bibitem{sawtooth}
Note that precisely this motion is predicted, on a global rather than
  mesoscopic scale, for perfectly ordered foams. See, \eg\
  Ref.~\protect\cite{Kraynik88}.
\bibitem{footflat}
This might allow `flat' moduli $\Gcomp(\om)$ ($x\approx 1$) to be found
  alongside a nonzero yield stress with power law flow exponent around 1/2
  ($x\approx 1/2$) ~\protect\cite{MasBibWei95,MasBibWei96,PriKis89}.
\end{thebibliography}
\end{document}